\documentclass[final,5p,times]{elsarticle}

\usepackage{amssymb}
\usepackage{amsmath}
\usepackage{lipsum}

\usepackage[colorlinks=true,linkcolor=blue]{hyperref}%
\usepackage{physics}
\usepackage{graphicx}
\usepackage{mwe}
\usepackage{caption}

\usepackage{dcolumn}
\usepackage{bm}
\usepackage{adjustbox}
\usepackage{rotating}
\usepackage{xcolor}
\usepackage{ulem}

\journal{Physics Letters B}
\usepackage{xcolor}

\begin{document}

\begin{frontmatter}



\title{Influence of the symmetry energy on the nuclear binding energies and the neutron drip line position}


\author[first]{Ante Ravli\'c}
\ead{aravlic@phy.hr}
\affiliation[first]{organization={Department of Physics, Faculty of Science, University of Zagreb},
            addressline={Bijenicka cesta 32}, 
            city={Zagreb},
            postcode={10000}, 
            country={Croatia}}
\author[second]{Esra Y\"uksel}
\ead{e.yuksel@surrey.ac.uk}
\author[first]{Tamara Nik\v{s}i\'c}
\ead{tniksic@phy.hr}
\author[first]{Nils Paar}
\ead{npaar@phy.hr}

\affiliation[second]{organization={Department of Physics, University of Surrey},
            city={ Guildford, Surrey},
            postcode={GU2 7XH}, 
            country={United Kingdom}}

\begin{abstract}
 
 A clear connection can be established between properties of nuclear matter and finite-nuclei observables, such as the correlation between the slope of the symmetry energy and dipole polarizability, or between compressibility and the isoscalar monopole giant resonance excitation energy. Establishing a connection between realistic atomic nuclei and an idealized infinite nuclear matter leads to a better understanding of underlying physical mechanisms that govern nuclear dynamics. In this work, we aim to study the dependence of the binding energies and related quantities (e.g. location of drip lines, the total number of bound even-even nuclei) on the symmetry energy $S_2(\rho)$. The properties of finite nuclei are calculated by employing the relativistic Hartree-Bogoliubov (RHB) model, assuming even-even axial and reflection symmetric nuclei. Calculations are performed by employing two families of relativistic energy density functionals (EDFs), based on different effective Lagrangians, constrained to a specific symmetry energy at saturation density $J$ within the interval of $30$--$36$ MeV. Nuclear binding energies and related quantities of bound nuclei are calculated between $8 \leq Z \leq 104$ from the two-proton to the two-neutron drip line. As the neutron drip line is approached, the interactions with stiffer $J$ tend to predict more bound nuclei, resulting in a systematic shift of the two-neutron drip line towards more neutron-rich nuclei. Consequentially, a correlation between the number of bound nuclei $N_{nucl}$ and $S_2(\rho)$ is established for a set of functionals constrained using the similar optimization procedures. The direction of the relationship between the number of bound nuclei and symmetry energy highly depends on the density under consideration.
\end{abstract}

\end{frontmatter}





In the nuclear landscape, the limits of stability are represented by the nuclear drip lines, which determine the maximum number of protons and neutrons within a nucleus before it decays. 
The accurate determination of the drip lines and properties of weakly bound nuclei is of utmost importance, not only from the standpoint of nuclear structure but also for astrophysics and the synthesis of chemical elements \cite{MUMPOWER201686}. Presently, it is known that more than half of nuclei heavier than iron are produced through the $r-$process, whose path is found near the neutron drip line \cite{THIELEMANN2011346,QIAN2007237}. The major challenge lies in the fact that, despite notable advancements in recent years \cite{Baumann_2012,BLANK2008403}, these nuclei remain beyond the reach of experiments.
While the proton drip line was experimentally uncovered up to protactinium ($Z = 91$) \cite{PhysRevLett.100.012501}, the neutron drip line is confirmed only up to $Z = 10$, and possibly $Z = 11$ \cite{PhysRevLett.123.212501,NOTANI200249,PhysRevLett.129.212502}. 
Hence, the accurate determination of nuclear properties and pinpointing the position of the neutron drip lines continue to heavily depend on theoretical calculations \cite{Erler2012,AFANASJEV2013680,MOLLER1988213,Ravlic2023}.


The nuclear energy density functional (EDF) theory represents an appropriate tool for studying the properties of finite nuclei across the nuclide chart, from the valley of stability up to the drip lines \cite{meng2016relativistic,colo_DFT}, as well as the nuclear equation of state (EOS) around the saturation densities.
Up to now, the nuclear EDFs have been mainly constrained by using the nuclear ground-state properties (such as masses, radii, and spin-orbit splitting). Although the isoscalar sector of EDFs is well constrained using the nuclear properties, their isovector component remains elusive and is the main source of uncertainty in the calculations \cite{PhysRevC.87.044320}. It has been known that the isovector part of the EDF determines the characteristics of neutron-rich drip line nuclei and plays an important role in determining the density dependence of the nuclear symmetry energy, which is a fundamental aspect of the nuclear EOS and remains beyond the realm of experimental measurement.
The nuclear symmetry energy contains information related to the proton-neutron asymmetry in the system, and its density dependence plays an important role in determining the properties of finite nuclei, especially near the drip lines \cite{Horowitz_2014,BALDO2016203}. It has also important applications in nuclear astrophysics, such as in the determination of the neutron star mass-radius relation \cite{Gandolfi2014,Lattimer2014}. In recent years, the detection of gravitational wave signatures from neutron star mergers has also opened up an entirely new avenue for constraining the symmetry energy through gravitational wave astronomy \cite{Zhang2019,PhysRevC.102.065805,Fattoyev2014}.
Therefore, there has been a great interest in studying the correlation between experimental observables and the nuclear EOS, which allows us to better constrain the EDF parameters and provides a simple means of understanding complex phenomena in finite-size nuclei. In particular, the correlations between the neutron skin thickness, dipole polarizability, and the symmetry energy around the saturation densities have been well established \cite{BALDO2016203, PhysRevLett.85.5296, PhysRevC.88.024316}. Using the experimental data on dipole polarizability in ${}^{208}$Pb for the first time in the optimization of a new relativistic functional, the symmetry energy (and its slope) around the saturation densities was obtained as 31.12 (46.32) MeV \cite{PhysRevC.99.034318}, in agreement with previous studies (see Ref. \cite{ROCAMAZA201896} and the references therein). However, the recent experimental data from the CREX \cite{PhysRevLett.129.042501} and PREX \cite{PhysRevLett.108.112502, PhysRevLett.126.172502} experiments could not provide a consistent solution to the symmetry energy problem, and its precise value remains as an unresolved issue for nuclear theory \cite{YUKSEL2023137622,PhysRevLett.129.232501}. While nuclear masses are known to constrain the isoscalar part of the EDFs, they can also provide insights into the symmetry energy parameters. However, the extent of their impact on the isovector part of the interaction remains unanswered. Considering that the symmetry energy is associated with the energy cost of the system as the neutron-to-proton ratio changes, it is expected to influence the binding energy of neutron-rich nuclei \cite{ROCAMAZA201896, CHEN2015284}. Therefore, further studies are needed to understand the nuclear symmetry energy and its impact on nuclear structure, especially nearby the drip lines. It is the aim of the present work to investigate the connection between the binding energies of atomic nuclei and the symmetry energy. To avoid any systematic dependence on a selected isotopic chain, we calculate the masses or binding energies of nuclei across the nuclide chart using relativistic EDFs with different underlying interactions. Then, we investigate the dependence of binding energies, the location of the neutron drip line, and the total number of bound nuclei on the symmetry energy around the saturation densities.

Since the majority of experiments in nuclear physics probe the properties of nuclei at or around nuclear saturation ($\rho_{0} \approx$ 0.15-0.16 fm$^{-3}$), it is customary to expand the nuclear EOS around the saturation densities and define it in terms of certain parameters \cite{ROCAMAZA201896}. The general expression for the energy per particle of nuclear matter is given with respect to the isospin-asymmetry parameter $\delta = (\rho_n - \rho_p) / (\rho_n + \rho_p)$ and neutron (proton) densities $\rho_{n(p)}$ \cite{BALDO2016203,ROCAMAZA201896}
\begin{equation}
E(\rho, \delta) = E(\rho, 0) + S_2(\rho) \delta^2 + \mathcal{O}(\delta^4).
\end{equation}
The first term corresponds to the energy of symmetric nuclear matter, while the second term is the symmetry energy as a function of density. Basically, the symmetry energy represents the difference in energies between pure neutron matter ($\delta=1$) and symmetric nuclear matter ($\delta=0$). It is customary to expand it as follows:
\begin{equation}\label{eq:S_2_expansion}
S_2(\rho) = J + L \left(\frac{\rho - \rho_0}{3 \rho_0}\right) + \frac{1}{2} K_{sym} \left( \frac{\rho - \rho_0}{3 \rho_0} \right)^2 + \mathcal{O}[(\rho-\rho_{0})^{3}].
\end{equation}
Here, $J$ represents the symmetry energy at saturation density, while $L$ represents its slope, and $K_{sym}$ is the incompressibility of the symmetry energy at saturation.

\begin{table}[h!]
\centering
\caption{The list of relativistic EDFs considered in this work, along with their nuclear matter properties at saturation density, including the incompressibility $K_0$, symmetry energy  $J$ and its slope $L$ \cite{PhysRevC.71.024312,universe7030071,PhysRevC.68.024310}.}\label{tab:tab1}
\begin{tabular}{cccc}
\hline
\hline
EDF & $K_0$ [MeV] & $J$ [MeV] & $L$ [MeV] \\
\hline
DD-PCJ30 & 230 & 30 & 35.6 \\
DD-PCJ32 & 230 & 32 & 52.3 \\
DD-PCJ34 & 230 & 34 & 72.1 \\
DD-PCJ36 & 230 & 36 & 94.1 \\
DD-PC1  & 230 & 33 & 70 \\
DD-PCX  & 213 & 31.1 & 46.3 \\
DD-MEJ30 & 250 & 30 & 29.7 \\
DD-MEJ32 & 250 & 32 & 46.4\\
DD-MEJ34 & 250 & 34 & 61.8\\
DD-MEJ36 & 250 & 36 & 84.8\\
DD-ME2 &  250 & 32.3 & 51.0 \\
\hline
\end{tabular}
\end{table}

As mentioned above, the symmetry energy and its density dependence play a significant role in determining the properties of neutron-rich nuclei. Consequently, they are expected to influence the position of the drip lines. However, establishing a correlation between the position of the neutron drip line and nuclear matter properties around the saturation densities remains puzzling \cite{PhysRevC.89.054320,Pearson2014,PhysRevC.103.034330}. A straightforward way toward understanding the underlying correlations involves constructing the EDFs with their parameters optimized for specific properties of nuclear matter. In this work, we study the impact of the symmetry energy parameters on the location of the drip lines using relativistic EDFs. To this aim, large-scale calculations are performed for even-even nuclei between $8 \leq Z \leq 104$ using the relativistic point-coupling DD-PC and density-dependent meson-exchange DD-ME family of functionals, optimized by imposing an additional constraint on the symmetry energy values at saturation densities as $J = 30, 32, 34, 36$ MeV \cite{universe7030071,PhysRevC.68.024310}.
The main advantage of these two families is that the same optimization protocols were used in constraining their parameters, with the same set of ground-state properties such as masses, charge radii, surface thickness and others. In addition to these EDFs with the constraint on $J$, we also employ the DD-PC1 \cite{PhysRevC.78.034318}, DD-PCX \cite{PhysRevC.99.034318}, and DD-ME2 \cite{PhysRevC.71.024312} functionals, where the model parameters were optimized by using a handful of nuclear ground-state observables in addition to those of nuclear matter (DD-PC1, DD-ME2), or more recently constrained using the nuclear collective excitation properties (DD-PCX). The EDFs employed in this work along with their nuclear matter parameters are given in Tab. \ref{tab:tab1}. All employed functionals use the same form of the separable pairing interaction \cite{PhysRevC.80.024313}, where pairing constants for neutrons(protons) $G_{p(n)}$ are included within the optimization procedure for DD-PCX and DD-PCJ family of EDFs. For the other functionals we assume set of parameters from Ref. \cite{PhysRevC.80.024313}. A total of 11 relativistic EDFs are employed in this work. We assume axial and reflection-invariant nuclear shapes with time-reversal symmetry. The resulting relativistic Hartree-Bogoliubov (RHB) equations are solved self-consistently by discretizing in the space of the axially-deformed harmonic oscillator, whose basis is defined by the number of oscillator shells $N_{osc}$ and oscillator length $b_0$ \cite{NIKSIC20141808}. We employ $N_{osc} = 20$, ensuring good convergence (see Supplementary material). 
In order to determine the ground state, we minimize the binding energy with respect to quadrupole deformation $\beta_2$. Such calculation is performed by the constrained RHB through the augmented Lagrangian method \cite{Staszczak2010}. We select 11 deformation mesh points within the interval $\beta_2 \in [-0.6, +0.6]$ and perform constrained RHB calculation for the first 20 iterations, after which the constraint is released. The total number of bound nuclei, denoted as $N_{nucl}$, is defined as the number of nuclei between the neutron and proton drip lines. In this work, we focus on even-even nuclei, which are considered bound if their two-neutron (proton) separation energy is $S_{2n(2p)} > 0$. Starting from the binding energy $E(Z,N)$ of a nucleus with $Z$ protons and $N$ neutrons, two-neutron and proton separation energies are calculated by
\begin{align}
S_{2n} &= E(Z,N-2) - E(Z, N), \\
S_{2p} &= E(Z-2,N) - E(Z,N),
\end{align}
where $E(Z,N) < 0$ by definition.

\begin{figure}[t!]
    \centering
    \includegraphics[width=\linewidth]{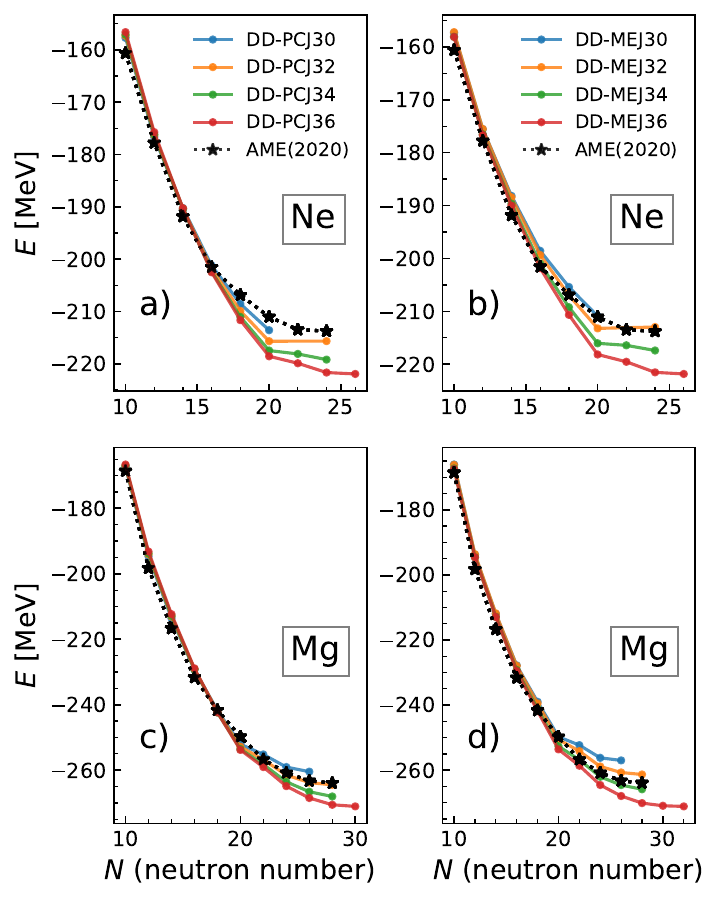}
    \caption{The isotopic dependence of binding energy $E$ for neon (a)--(b) and magnesium (c)--(d) isotopic chains. Calculations are performed with DD-PCJ (a,c) and DD-MEJ (b,d) effective interactions. Experimental data is taken from Ref. \cite{nndc}.}
    \label{fig:Ne_and_Mg}
\end{figure}

Based on the liquid-drop (LD) formula, it is easy to deduce that symmetry energy parameter $a_{sym}$ directly impacts the neutron drip line, and is related to $J$ \cite{ROCAMAZA201896,BALDO2016203}. Increasing the value of $a_{sym}$ leads to nuclei preferring to be closer to the valley of stability, thus making the neutron drip line less neutron-rich \cite{PhysRevC.82.027301}.
In Ref. \cite{PhysRevC.89.054320}, a large-scale calculation of nuclear landscape for even-even nuclei was performed using a set of relativistic EDFs. A possibility of relating the location of a two-neutron drip line with symmetry energy at saturation density $J$ was investigated, however, no correlation was found. On the other hand, in Ref. \cite{PhysRevC.92.031303}, a negative correlation between the number of bound nuclei and the symmetry energy at sub-saturation densities of $0.11/0.16 \times \rho_0$ was established through the utilization of 42 relativistic and non-relativistic EDFs. It is interesting to study such seemingly contradictory conclusions stemming from different theoretical models. We show that all conclusions can indeed be correct by extending our considerations to the full density-dependence of the symmetry energy $S_2(\rho)$. Since it is difficult to define the average nuclear density, we avoid this notion altogether, and mainly focus on densities at and  below the saturation.


The influence of increasing symmetry energy on the characteristics of finite nuclei has already been studied in Ref. \cite{CHEN2015284}, for oxygen and calcium isotopic chains, in which the authors used the same family of functionals differing only in the symmetry energy value. It has been found that a larger or stiffer nuclear symmetry energy around saturation densities favors neutron-rich systems more by increasing the total binding energy of the system. Consequently, it predicts a delay in reaching the neutron drip line compared to softer ones. Although the conclusions are supported by a limited set of existing experimental data, it is important to generalize the considerations by performing a study across the nuclide chart, with various parametrizations of underlying model Lagrangians.

In this work, we aim to address the sensitivity of the neutron drip line to nuclear symmetry energy by performing extensive calculations with 11 EDFs for the whole nuclide chart. First, we want to examine the behavior of nuclei when the symmetry energy of nuclear matter is increased. Our attention is focused on neutron-rich side of the nuclear landscape since the effect of symmetry energy on proton-rich nuclei is reduced by the Coulomb barrier. To achieve this, we employ the families of relativistic EDFs specifically constrained to a given value of symmetry energy at saturation $J$. In this way, we can systematically control $J$ and study its influence on specific nuclear properties. To benchmark our calculations with experimental data, we have only a few even-even isotopic chains, since the neutron drip line is presently confirmed up to sodium ($Z = 11$) \cite{PhysRevLett.129.212502}. Since the analysis for oxygen was performed in Ref. \cite{CHEN2015284}, here we focus on the following lightest isotopic chains: neon ($Z = 10$) and magnesium ($Z = 12$). In Figure \ref{fig:Ne_and_Mg}, the results are shown for the total binding energy $E$ for the isotopic chains of neon (a)--(b) and magnesium (c)--(d). The results are displayed for both DD-PCJ and DD-MEJ families of functionals with $J = 30,32,34$ and 36 MeV. Calculations are performed up to the two-neutron drip line. For both chains and effective interactions, we observe a similar behavior with increasing neutron number. The effect of increasing $J$ on the binding energy is negligible for nuclei closer to the valley of stability, meaning that they are less sensitive to the isovector properties of underlying EDFs. However, as the two-neutron drip line is approached, a clear separation of binding energy curves among different functionals is observed. In general, increasing $J$ makes nuclei more bound, with a more pronounced effect towards the neutron drip line. By increasing the value of $J$, not only does the binding energy increase, but so does its slope, thus the position of the two-neutron drip lines is also affected. In other words, in a family of relativistic nuclear EDFs, a functional with stiffer symmetry energy predicts a delay in reaching the neutron drip lines compared to a softer one (see Supplementary material for other isotopic chains). 
It is observed that experimental data prefers softer symmetry energy values ($J = 30, 32$ MeV). The neutron drip line is reached for ${}^{34}$Ne, consistent with $J = 32$ and 34 MeV, for both effective interactions. Recently, the near drip line nucleus ${}^{40}$Mg was observed \cite{PhysRevLett.122.052501}, although the exact location of the drip line for the magnesium isotopic chain has not yet been confirmed. 
However, we can still test how our calculations compare to the experiment. Results are shown in Fig. \ref{fig:Ne_and_Mg}(c)--(d) where the ${}^{40}$Mg is predicted to be a drip line nucleus for $J = 32, 34$ MeV for both effective interactions, while it is not bound for $J = 30$ MeV. Based on the experimental data for Ne and Mg chains, we are able to exclude $J = 30$ MeV as under-predicting the location of the neutron drip line. On the other hand, $J = 36$ MeV is most likely over-predicting its location in these two chains. However, as the neutron drip line is further uncovered with the development of modern experimental facilities, systematic comparison of its location with theoretical calculations could lead to better constraints on $J$.

\begin{figure}[t!]
\centering
\includegraphics[scale = 0.32]{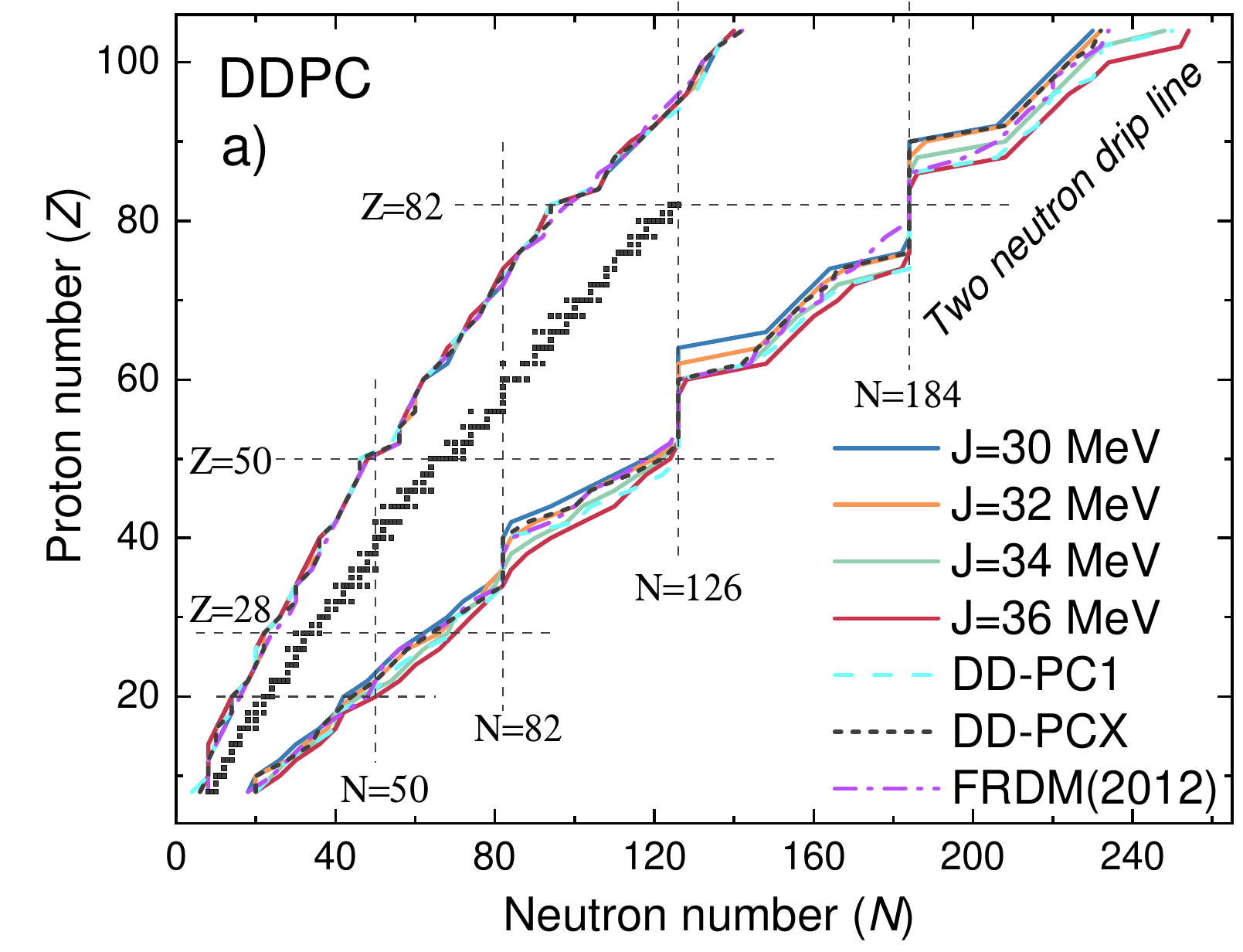}
\includegraphics[scale = 0.32]{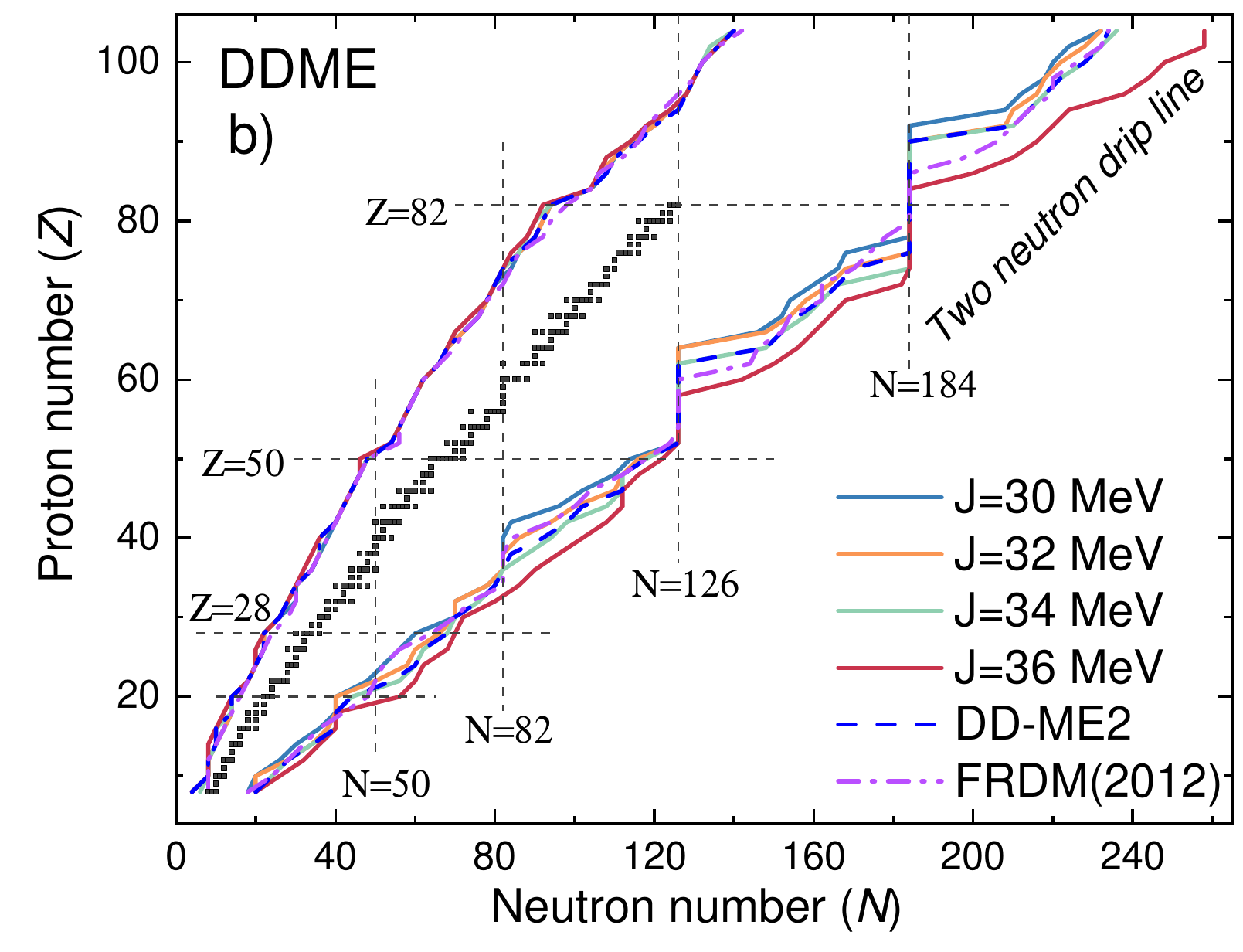}
\caption{Two-proton and two-neutron drip lines for nuclei between $8 \leq Z \leq 104$. The calculations are performed within the axially-deformed RHB using the DD-PCJ (a) and DD-MEJ (b) families of relativistic EDFs for $J = 30$--36 MeV. The results are also presented for the DD-PC1 and DD-ME2 functionals as well as the FRDM(2012) \cite{MOLLER20161}. The dashed lines denote the shell closure numbers. The black squares represent the stable nuclei 
 \cite{nndc}.}\label{fig:drip_dd}
\end{figure}

On a larger scale, the effect of the increasing symmetry energy at saturation density $J$ on two-nucleon drip lines is shown in Fig. \ref{fig:drip_dd}(a)--(b). In Fig. \ref{fig:drip_dd}(a), we present the results for the DD-PCJ family of functionals as well as DD-PC1, DD-PCX, while in Fig. \ref{fig:drip_dd}(b), we show the results for the DD-MEJ functionals and DD-ME2. We have also included the results from FRDM(2012) \cite{MOLLER20161} for the purpose of comparison. Both families of functionals predict similar trends on drip lines. The results from the FRDM(2012) are also in agreement with them. First, as expected, the effect of increasing $J$ on the proton drip line is negligible, and all functionals predict similar results. On the other hand, the two-neutron drip line shows a systematic shift towards a higher neutron number with increasing $J$. It is interesting to notice that increasing $J$ tends to push the neutron drip line beyond the $N = 50$ and $N = 82$ shell closures, whereas the signature of $N = 126$ and $N = 184$ shell closures is still clearly seen up to $J = 36$ MeV. Similar trends are also observed between DD-PC1 and DD-PCX, where the latter has a lower $J$ value and reaches the drip line earlier than the former one. Based on these results, we should expect an increasing number of bound nuclei with increasing $J$. However, as we will discuss in the following, such conclusions can only be drawn for functionals with systematic variation of $J$. In general, weaker trends are found for other functionals \cite{AFANASJEV2013680}.

\begin{figure}[t!]
\centering
\includegraphics[scale = 0.7]{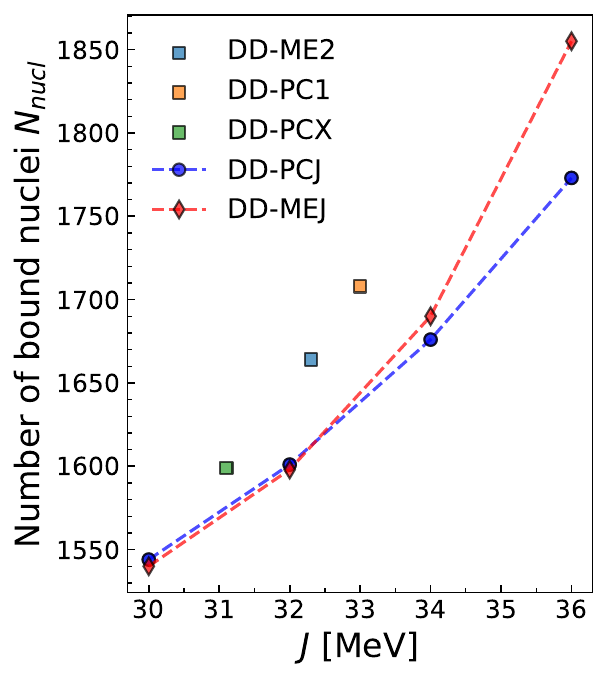}
\caption{The number of bound even-even nuclei $N_{nucl}$ as a function of the symmetry energy at saturation density $J$. The calculations are performed with 11 different relativistic EDFs.}\label{fig:N_nucl_J}
\end{figure}

In Figure \ref{fig:N_nucl_J}, we show the total number of bound even-even nuclei $N_{nucl}$ for 11 relativistic EDFs with respect to the symmetry energy at saturation density $J$ of a given functional. Blue and red dashed lines connect results for the EDFs constrained to a specific $J$ value (30, 32, 34, and 36 MeV), while others (DD-PC1, DD-PCX, and DD-ME2) are denoted as squares. Naively, based on these calculations, one could claim that the number of nuclei increases with $J$; however, as we will demonstrate later, such statements can only be made with functionals constrained systematically by using the same optimization protocols (the same set of ground- and excited-state observables for a similar collection of nuclei). It is important to remember that $J$-dependence corresponds only to a zeroth-order expansion of $S_2(\rho)$ [cf. Eq. (\ref{eq:S_2_expansion})] and does not explore the full density dependence of the symmetry energy.

\begin{figure}[t!]
\centering
\includegraphics[width = \linewidth]{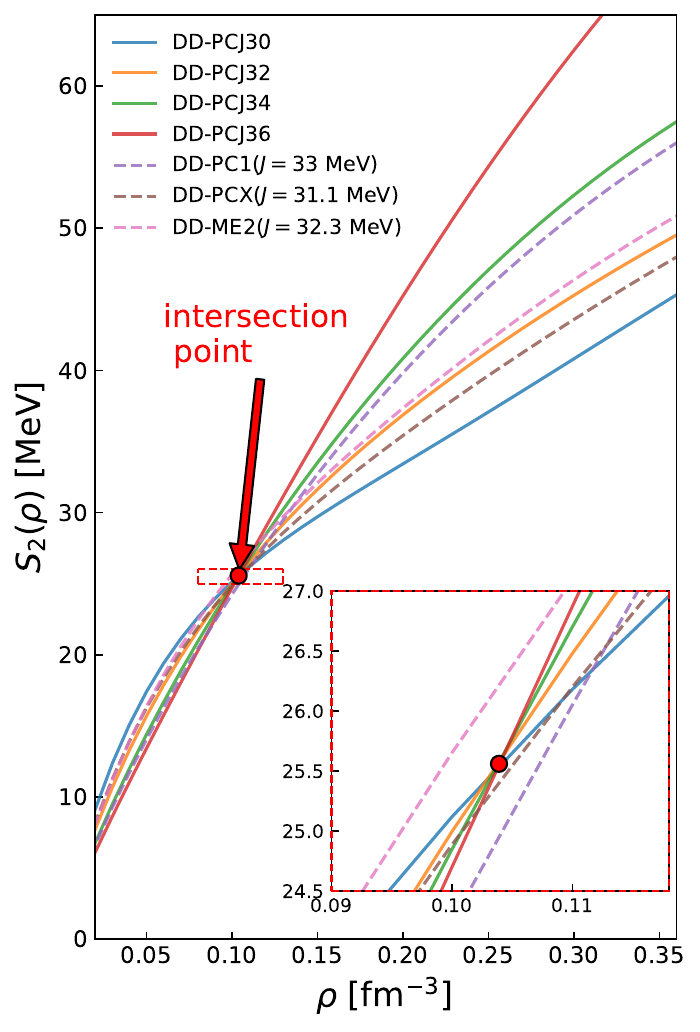}
\caption{Density-dependence of the symmetry energy $S_2(\rho)$ calculated with relativistic point-coupling EDFs, including the DD-PCJ family (solid lines), as well as DD-PC1, DD-PCX, and DD-ME2 (dashed lines). The inserted figure shows an enlarged region in the vicinity of the intersection point (red dashed rectangle). The point where $S_2$ of different DD-PCJ functionals intersect is labeled with a red dot.}\label{fig:S_2}
\end{figure}

To infer a full density dependence of the symmetry energy, in Fig. \ref{fig:S_2}, we plot the symmetry energy as a function of the density for the set of relativistic EDFs. One can observe that the behaviour of the symmetry energy is quite diverse and uncertain above the saturation density, which is a long-standing problem due to the lack of the appropriate constraints. However, we are dealing with finite nuclei, thus focusing on the region around the saturation density. The calculations of $S_2$ reveal a clear hierarchy among the EDFs and follow an increasing trend with density. For the family of DD-PCJ functionals, one can observe a point where the $S_2$ curves intersect. In the following, we will refer to this point of intersection and label its density as $\rho_{c}$. In the case of the shown DD-PCJ EDFs, $\rho_c \sim 0.10$ fm${}^{-3}$. For densities below intersection, $\rho < \rho_c$, the symmetry energy tends to decrease with increasing $J$ of the EDF, while for densities above  $\rho > \rho_c$, the opposite is true. Note that $\rho_c$ lies below the saturation density $\rho_0$, while it is close to the so-called average nuclear density \cite{PhysRevC.92.031303}. The inserted panel on Fig. \ref{fig:S_2} shows an enlarged region near the $\rho_c$. It can be noted that the DD-PCJ family of functionals indeed intersects at $\rho_c$, while the DD-PC1, DD-ME2 and even DD-PCX visibly depart from this point. Unlike the DD-PCJ family (or even DD-PCX), the DD-PC1 and DD-ME2 EDFs were optimized using different protocols \cite{PhysRevC.78.034318,PhysRevC.71.024312}.

\begin{figure}[t!]
\centering
\includegraphics[scale = 0.8]{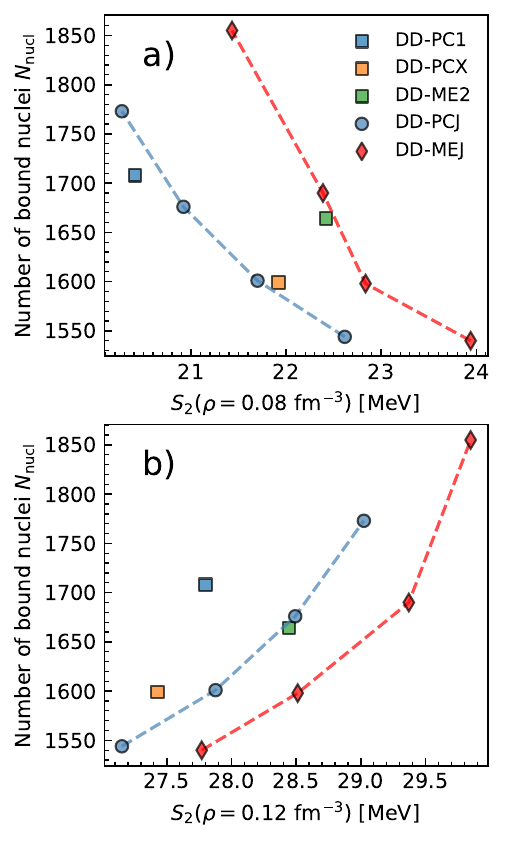}
\caption{The total number of bound nuclei $N_{nucl}$ shown as a function of the symmetry energy $S_2$ at $\rho = 0.08$ fm${}^{-3}$ (a) and $\rho = 0.12$ fm${}^{-3}$ (b). Results for DD-PCJ (blue) and DD-MEJ families of EDFs are shown with dashed lines, together with results for DD-PC1, DD-PCX and DD-ME2 functionals (squares).}\label{fig:Nnucl_S_2_dependence}
\end{figure}

In the following, we select two density points, one below the intersection density $\rho = 0.08 \textrm{ fm}^{-3} < \rho_c$, and the other above, $\rho = 0.12 \textrm{ fm}^{-3} > \rho_c$. In Figure \ref{fig:Nnucl_S_2_dependence}, we show $N_{nucl}$ for 11 relativistic EDFs with $S_2(0.08 \textrm{ fm}^{-3})$ (a) and $S_2(0.12 \textrm{ fm}^{-3})$ (b).
First, we focus our discussion on EDFs constrained to a specific $J$ (DD-PCJ and DD-MEJ families). For $\rho = 0.08$ fm${}^{-3}$ in Fig. \ref{fig:Nnucl_S_2_dependence}(a) we observe a decreasing trend in the number of bound nuclei with increasing symmetry energy $S_2$. On the other hand, at $\rho = 0.12$ fm${}^{-3}$ in Fig. \ref{fig:Nnucl_S_2_dependence}(b), the opposite trend is observed \textit{i.e.} number of bound nuclei increases with symmetry energy. It is interesting to note that results for DD-PCJ and DD-MEJ families lie on different lines describing the evolution with symmetry energy. Due to different underlying interactions, the DD-PCJ EDFs, for a fixed $J$, have lower value of symmetry energy $S_2$. Therefore, we see that studying correlation in the number of bound nuclei with a choice of a single value of symmetry energy around the saturation density provides an incomplete information and can infer misleading conclusion. In the case shown in Fig. \ref{fig:N_nucl_J}, density point at which $S_2(\rho)\sim J$ was located above the intersection point $\rho_c$, is leading to an increasing trend in the number of bound nuclei with $J$. In general, studying the full density-dependence of the symmetry energy allows for a complete picture, where $N_{nucl}$ can either increase or decrease with $J$.

However, such clear trends with symmetry energy can only be deduced for a family of functionals using similar optimization protocols and nuclear matter properties. Looking at results for DD-PC1, DD-PCX and DD-ME2 in Fig. \ref{fig:Nnucl_S_2_dependence}(a)--(b) we observe no clear trends. Before drawing conclusions, we should keep in mind that these EDFs use very different optimization procedures to determine their parameters. The DD-ME2 interaction is optimized using a handful of spherical nuclei and their ground-state properties \cite{PhysRevC.71.024312}. This is contrary to the DD-PC1 interaction, which used ground-state experimental data of axially-deformed nuclei. In the optimization of the DD-PCX interaction, information from excited states was also employed in addition to ground-state observables \cite{PhysRevC.99.034318}. Such a plethora of information is best reflected when trying to understand the behavior of symmetry energy for these three EDFs. The inserted panel in Fig. \ref{fig:S_2} shows an enlarged density region in the vicinity of the previously discussed intersection point $\rho_c \approx 0.10$ fm${}^{-3}$. Unlike DD-PCJ family, it is important to note that there is no single intersection point for these three functionals. First, at $\rho_c$ or slightly below, we observe that $S_2(\textrm{DD-ME2}) > S_2(\textrm{DD-PCX}) > S_2(\textrm{DD-PC1})$, which is reflected in Fig. \ref{fig:Nnucl_S_2_dependence}(a). Indeed, these EDFs do not follow the same trend of decreasing number of bound nuclei with symmetry energy. At $\rho = 0.12$ fm${}^{-3}$ and above, we have $S_2(\textrm{DD-ME2}) > S_2(\textrm{DD-PC1}) > S_2(\textrm{DD-PCX})$, and the order of DD-PC1 and DD-PCX is reversed, as confirmed in Fig. \ref{fig:Nnucl_S_2_dependence}(b). On a simple example of these three functionals, we demonstrated why it is difficult to infer any correlation between number of bound nuclei, i.e., the location of the drip lines, and symmetry energy by employing functionals with considerably different optimization protocols. These correlations can be assessed by considering functionals constrained in a systematic way for the range of values of symmetry energy, as discussed above.

In summary, in this Letter we have investigated how the binding energies and related quantities, the location of the drip lines, and
total number of bound even-even nuclei $N_{nucl}$ depend on the symmetry energy. Due to the density dependence of the symmetry energy, the results qualitatively differ if the density considered is located below or above the intersection point.
It is shown that the total number of bound nuclei exhibits either an increasing or decreasing trend with the symmetry energy depending on the value of density considered in the calculations. For general EDFs, it is misleading to establish a clear correlation between the position of the neutron drip line and symmetry energy. In order to show that such a correlation indeed exists, we employed specialized families of relativistic EDFs which are systematically constrained to a specific value of the symmetry energy at saturation density $J$. With those sets of EDFs, which use the same nuclear observables in their optimization procedure, it is possible to investigate the correlation between the location of the two-neutron drip line (and $N_{nucl}$) with the symmetry energy. This study could be further extended in future studies by also considering odd nuclei.

\section*{Acknowledgements}
This work is supported by the QuantiXLie Centre of Excellence, a project co financed by the Croatian Government and European Union through the European Regional Development Fund, the Competitiveness and Cohesion Operational Programme (KK.01.1.1.01.0004) (A.R., T.N., E.Y., N.P.). Support from the Science and Technology Facilities Council (UK) through grant ST/Y000013/1 is also acknowledged (E.Y.). This work was supported in part through computational resources and services provided by the Institute for Cyber-Enabled Research at Michigan State University (A.R.). We acknowledge support by the US National Science Foundation under Grant PHY-1927130 (AccelNet-WOU: International Research Network for Nuclear Astrophysics [IReNA])(A.R., N.P.).

\bibliographystyle{elsarticle-num} 
\bibliography{main}






\end{document}